\documentclass[11pt]{article}
\usepackage{amsthm,amsfonts,amssymb,amsmath}

\theoremstyle{plain}
\newtheorem{teo}{Theorem}

\newtheorem*{pro}{Proposition}
\newtheorem*{cor}{Corollary}
\theoremstyle{remark}
\newtheorem*{rem}{Remark}

\newcommand{\C}{{\mathbb C}}
\newcommand{\Z}{{\mathbb Z}}
\newcommand{\ie}{i.\,e.\ }
\newcommand{\ds}{\displaystyle}

\newcommand{\wh}{\widehat}

\newcommand{\p}{\partial}
\newcommand{\intl}{\int\limits}
\newcommand{\suml}{\sum\limits}

\begin{document}

\title{The dynamics of zeros of the elliptic solutions to the 
Schr\"odinger equation}
\author{A. Akhmetshin${}^*$\and Y. Volvovsky${}^{\dagger}$}
\date{}  
\maketitle

\begin{center}    
{\sffamily Columbia University}\\
{\small
2990 Broadway, Mailcode 4406, New York, NY 10027, USA}\\
and\\ 
{\sffamily Landau Institute for Theoretical Physics}\\
{\small Kosygina str.~2, Moscow 117334, Russia.}\\
\end{center}

\begin{center}
e-mail:\\
${}^*$\verb"alakhm@math.columbia.edu"\\
${}^{\dagger}$\verb"yurik@math.columbia.edu"
\end{center}
\medskip

\begin{abstract}
J.\,F.~van Diejen and H.\,Puschmann have recently shown 
that the dynamics of zeros of the $n$-solitonic solutions 
to the Schrodinger equation with the reflectionless
potential is governed by a rational Ruijsenaars\,--\,Schneider
system. We use the algebraic-geometrical construction of
solutions to the Schrodinger equation to generalize this 
result to the elliptic case.
\end{abstract}

\section{Introduction}

In the recent paper \cite{dp} it was noticed that the dynamics 
of zeros of $n$-solitonic solutions to the Schr\"odinger equation 
with the reflectionless potential is governed by the rational
Ruij\-senaars\,--\,Schneider system with the harmonic term \cite{rs}. 
This result appears to be surprising since the aforementioned
dynamics was described long ago, though in a different form.
In~\cite{dub} it was shown that the Bloch solution to the 
Schr\"odinger equation 
$$
(\p^2_x-u(x))\psi(x,E)=E\psi(x,E)
$$
with the finite-gap potential $u(x)$ is a well-defined function on 
the hyperelliptic curve
$$
y^{2}=\prod_{i=1}^{2g+1} (E-E_i).
$$ 
The zeros of this function satisfy the Dubrovin equations \cite{dub2}:
$$
\frac{\p \wh\gamma_s}{\p x}=\frac{2 y(\gamma_s)}{\prod_{j\ne s} 
(\wh\gamma_s-\wh\gamma_j)},
$$
where by $\wh\gamma$ we denote the projection of a point $\gamma$
on the $E$-plane. Notice that these equations contain parameters 
of the curve. An analog of the Dubrovin equations holds also
for degenerate hyperelliptic curves (these are described by the 
same equation where not all $E_{i}$'s are distinct) and in particular
for fully degenerate hyperelliptic curves which can be thought 
of as a Riemann sphere with $n$ couples of pairwise identified points. 
As it was shown in \cite{dp} in the latter case parameters of the 
curve can be excluded from the system. The modified system then is a
system of second-order differential equations written solely in 
terms of zeros of the corresponding function. It coincides with 
the Ruijsenaars-Schneider system and therefore is Hamiltonian, the
expressions for the parameters of the curve being the integrals of
motion.

In this paper we exploit the algebraic-geometrical approach 
developed in \cite{kr} to apply these ideas to the case of 
the reflectionless potentials on a background of finite zone 
potentials, corresponding to the elliptic curves with 
self-intersections.  
The dynamics of zeros of the corresponding solutions 
resembles the elliptic Ruijsenaars\,--\,Schneider system \cite{rs}.
We show that the system describing these dynamics is Hamiltonian 
and completely integrable, the angle-type variables being the 
analogs of the components of the Abel map.

We hope to come up with the general system describing the case of
the hyperelliptic curve with arbitrary degree of degeneracy shortly.

\section{Algebraic-geometrical data}

In this section we present some basic facts from the finite-gap 
theory. 

Consider an elliptic curve 
$\Gamma$, given by the equation 
\begin{equation}\label{cur}
y^2=E^{3}-g_{2}E-g_{3}.
\end{equation}
It's compactified at infinity by one point which we denote by 
$\infty$. The only (up to multiplication by constant) holomorphic 
differential on $\Gamma$ has the following form:
$\ds \omega^h=\frac{\raisebox{-1pt}{$dE$}}{\raisebox{2pt}{$y$}}$.
It defines the map from $\Gamma$ to the torus 
$\wh\Gamma=\C/\Z[2\omega_1,2\omega_2]$, where $2\omega_1$ 
and $2\omega_2$ are $a$- and $b$-periods of $\omega^h$, respectively. 
This map, given by
$$
A\colon P\longmapsto z=\int_{\infty}^P \omega^h  
$$
and known as the Abel map, allows us to identify $\Gamma$ and
$\wh\Gamma$. 

Corresponding to the torus $\wh\Gamma$ are the standard
Weierstrass functions 
$$\sigma(z|\omega_1,\omega_2),\quad 
\zeta(z|\omega_1,\omega_2)=\frac{\sigma'(z|\omega_1,\omega_2)}
{\sigma(z|\omega_1,\omega_2)},\quad \wp(z|\omega_1,\omega_2)=
-\zeta'(z|\omega_1,\omega_2)
$$ (see \cite{bat} for reference).
The function $\sigma(z)$ has the following properties:

i) in the neighborhood of zero $\sigma(z)=z+O(z^{5});$

ii) $\sigma(z+2\omega_j)=e^{2\eta_j(z+\omega_j)}\sigma(z)$, where
$\eta_j=\zeta(\omega_j)$.

Notice that $\wp(z)$ is an elliptic function with the only 
(double) pole at $z=0$ and $\zeta(z)$ has the simple pole 
at $z=0$ and satisfies the following monodromy conditions:
$$
\zeta(z+2\omega_1)=\zeta(z)+\eta_1,\quad 
\zeta(z+2\omega_2)=\zeta(z)+\eta_2.
$$

The map $z\mapsto (E=\wp(z),y=\wp'(z))$ is inverse to the Abel map.

Let us fix $n-1$ points $\kappa_{1},\dots,\kappa_{n-1}$ 
on~$\wh\Gamma$.

\begin{pro} \emph{\cite{kr1}}\ 
For generic divisor $D=\gamma_1+\dots+\gamma_n$ on the curve 
$\wh\Gamma$ there exists a unique function  $\psi(x, z|D)$ 
satisfying the following conditions:

1. It's  meromorphic on the curve $\wh\Gamma$ outside the point $z=0$ 
and has poles of at most first order at the points $\gamma_i$, 
$i=1,\dots,n$. 

2. In the neigborhood of $z=0$ it has a form
$$
\psi(x, z)=e^{x z^{-1}}(1+\sum_{s=1}^{\infty}\xi_{s}(x) z^{s}).
$$

3. $\psi(x,\kappa_{i})=\psi(x,-\kappa_{i}).$
\end{pro}
\begin{rem}
In general the function $\psi(x,z|D)$ is defined on the curve 
$\Gamma$ itself, but here for the sake of brevity we use the
identification between $\Gamma$ and $\wh\Gamma$.
\end{rem}

\begin{proof}
 The uniqueness of such a function follows immediately from
 the Riemann\,--\,Roch Theorem. To show the existence we shall
 consider the following function
\begin{equation}\label{psi}
\psi(x, z|D)=e^{\zeta(z) x}
\frac{\prod_{i=1}^{n}\sigma(z-z_{i}(x))}%
{\prod_{s=1}^n\sigma(z-\gamma_s) \prod_{i=1}^n\sigma(z_i(x))}.
\end{equation}  

The set of conditions $\psi(x,\kappa_{i})=\psi(x,-\kappa_{i})$ 
and the constraint $\sum_{i=1}^{n}z_{i}(x)=x$ (the latter means 
that~$\psi$ is an elliptic function) form the system of $n$ 
equations on the functions $z_{i}(x).$ For generic data this system 
is non-degenerate. Then it has the only solution (up to the 
permutations) and therefore defines the function $\psi(x, z|D)$ 
uniquely.
\end{proof}  
 
\begin{cor}
The above-constructed function $\psi(x, z|D)$ is a solution to 
the Schr\"odinger equation
\begin{equation}\label{sch}
(\partial_{x}^{2}+u(x))\psi(x, z)=\wp(z)\psi(x, z),
\end{equation}	 
where $u(x)=-2\sum_{i=1}^{n}\wp(z_{i}(x)) z'_i(x)$.
\end{cor}   
 
\begin{proof}
Consider a function 
$\psi_0(x, z)=(\partial_{x}^{2}+u(x)-\wp(z))\psi(x, z)$. 
It's straightforward to check that the function $\psi+\psi_0$ 
satisfy all defining properties of the function $\psi$. 
The uniqueness of $\psi$ implies that $\psi_0=0$.
\end{proof}

\section{Main results}

\begin{teo}
The zeros of the function $\psi(x,z|D)$ satisfy the following 
dynamics:
\begin{equation}\label{soe}
z_{i}''=\sum_{k\ne i}z_{i}'z_{k}'
\frac{\wp'(z_{i})+\wp'(z_{k})}{\wp(z_{i})-\wp(z_{k})},
\quad i=1,\dots,n .
\end{equation}
\end{teo} 

\begin{proof}
To obtain these equations one has to divide (\ref{sch}) by 
$\psi(x, z)$ and compare the residues of the both sides of
the obtained equation at the points $z_{i}(x).$
\end{proof}
 
\begin{rem}
Theorem 1 provides us with a wide class of solutions to system 
(\ref{soe}) coming from the algebraic-geometrical data. The simple
"dimensional" argument shows that in fact these are all solutions.
We could reverse the whole reasoning starting with the solution
to (\ref{soe}) and showing that the corresponding elliptic function
(\ref{psi}) solves the Shr\"odinger equation.
\end{rem}
    
From now on we shall study system (\ref{soe}). 
Let us introduce the variables $\xi_i=\ln z'_i$, $i=1,\dots,n$.  
In the variables $z_i$, $\xi_i$ system (\ref{soe}) has 
the following form:
\begin{equation}\label{sxi}
\begin{aligned}
z'_i&=e^{\xi_i},\\
\xi'_i&=\sum_{k\ne i} e^{\xi_k}
\frac{\wp'(z_{i})+\wp'(z_{k})}{\wp(z_{i})-\wp(z_{k})},
\quad i=1,\dots,n .
\end{aligned}
\end{equation}

\begin{pro}
System (\ref{sxi}) is Hamiltonian with respect to the Hamiltonian 
$\ds H=\sum_{i=1}^n e^{\xi_j}$ and a 2-form
\begin{equation}\label{o1}
\omega=\sum_{i=1}^n dz_i\wedge d\xi_i-\frac{1}{2}\sum_{i\ne j}
\frac{\wp'(z_{i})+\wp'(z_{j})}{\wp(z_{i})-\wp(z_{j})}
dz_i\wedge dz_j .
\end{equation}
\end{pro}

The proof is a straigtforward calculation.    

Note that 
$$
\begin{aligned}
\omega &=\sum_{i=1}^n dz_i\wedge d\xi_i-\sum_{j\ne i}
\frac{\wp'(z_{i})}{\wp(z_{i})-\wp(z_{j})} dz_i\wedge dz_j = {}\\
{}&=\sum_{i=1}^n dz_i\wedge d\xi_i+\sum_{j\ne i} dz_i\wedge
\frac{\wp'(z_j)\,dz_j-\wp'(z_i)\,dz_i}{\wp(z_j)-\wp(z_i)}={}\\
{}&=\sum_{i=1}^n dz_i\wedge d\xi_i+\sum_{i\ne j}dz_i\wedge
d\Bigl(\ln(\wp(z_j)-\wp(z_i))\Bigr)=
\sum_{i=1}^n dz_i\wedge d\rho_i ,
\end{aligned}
$$
where 
$$
\rho_i=\xi_i+\sum_{j\ne i}\ln(\wp(z_j)-\wp(z_i)) .
$$

The algebraic-geometrical construction from the previous section 
provides us with a hint on how the first integrals of system 
(\ref{soe}) should look like.
The constraints $\psi(x,\kappa_{s})=\psi(x,-\kappa_{s})$ imply
the equations 
$$
\sum_{j=1}^{n}\frac{z_{j}'}{\wp(\kappa_{s})-\wp(z_{j})}=0 ,
$$
which can be rewritten in the following form
$$
\sum_{i=1}^n z'_i\prod_{j\ne i} (\wp(z_j)-\wp(\kappa_s))=0 .
$$
These considerations motivate 

\begin{teo}
The coefficients $H_k$ of the polynomial 
\begin{equation}\label{l}
L(\lambda|z,z')=\sum_{k=0}^{n-1} H_k(z,z')\lambda^k=
\sum_{i=1}^n z'_i \prod_{j\ne i} (\wp(z_j)-\lambda)
\end{equation}
are the integrals of motion of system (\ref{soe}).
\end{teo}

\begin{rem} 
Note that the leading cofficient $H_{n-1}(z,z')$ of $L$ 
is equal up to the sign to the Hamiltonian $H(z,z')$ of 
system (\ref{soe}). 
\end{rem}

The statement of the theorem is clear since we know that 
all solutions are algebraic-geometrical. 
However, we would like to present an independent 
direct proof. It can be found in the Appendix I. 

Let us notice that $L(\wp(z_j))=e^{\rho_j}$.
Using this identity we can rewrite the form $\omega$
in the following way:
$$
\begin{aligned}
\omega &=\sum_{i=1}^n dz_i\wedge d\rho_i=
\sum_{i=1}^n dz_i\wedge d\ln L(\wp(z_i))={}\\
{}&=\sum_{i=1}^n \frac{1}{L(\wp(z_i))}\,dz_i\wedge
d\left( \sum_{s=0}^{n-1} H_s \wp^s(z_i) \right)=
\sum_{i=1}^n\sum_{s=0}^{n-1}\frac{\wp^s(z_i)}{L(\wp(z_i))}
\,dz_i\wedge dH_s={}\\
{}&=
\sum_{s=0}^{n-1} d\left(\sum_{i=1}^n \intl^{\wp(z_i)}
\frac{E^s\,dE}{L(E)y(E)} \right)\wedge dH_s+
\sum_{s,k=0}^{n-1} \left(\sum_{i=1}^n \intl^{\wp(z_i)}
\frac{E^{s+k}\,dE}{L(E)^2y(E)} \right)dH_k\wedge dH_s={}\\
{}&=
\sum_{s=0}^{n-1} d\left(\sum_{i=1}^n \intl^{\wp(z_i)}
\frac{E^s\,dE}{L(E)y(E)} \right)\wedge dH_s ,
\end{aligned}
$$
where the function $y(E)$ is given by (\ref{cur}).

Thus we have proved the following statement.

\begin{teo}
The variables 
$$
\varphi_s=\sum_{i=1}^n \intl^{\wp(z_i)}
\frac{E^s\,dE}{L(E)y(E)},\qquad s=0,\dots,n-1
$$ 
and $H_s$ defined by (\ref{l}) are the action-angle type 
variables for system (\ref{soe}).
\end{teo}    

We would like however to rewrite the form $\omega$ once 
again in terms of the zeros of the polynomial 
$L(\lambda|z,z')$ which we shall denote by~$\wh\kappa_j$, 
$j=1,\dots,n-1$. In order to do this we introduce the 
new variables 
$$
\chi_j=\sum_{i=1}^n \intl^{\wp(z_i)}
\frac{dE}{(E-\wh\kappa_j)y(E)},\qquad j=1,\dots,n-1 .
$$ 
Let us also introduce the variable 
$$
\chi=\sum_{i=1}^n \intl^{\wp(z_i)}\frac{dE}{y(E)} .
$$

\begin{teo}
The above-defined form $\omega$ admits the following 
representation
\begin{equation}\label{ok}
\omega=d\chi\wedge d(\ln H)+\sum_{j=1}^{n-1} d\chi_j\wedge 
d\wh\kappa_j.
\end{equation}
\end{teo}

\begin{rem} 
We want to emphasize the fact that the variables
$\{\chi,\chi_j,\,j=1,\dots,n-1\}$ are the degenerate curve
analogs of the components of the Abel map. So our Hamiltonian 
structure fits in the general scheme proposed in {\cite{ves}}
and developed in \cite{kp}.
\end{rem}

The proof is a straightforward computation (see Appendix II).

\section*{Appendix I}
Let us consider the polynomial 
$$
L(\lambda|z,z')=\sum_{j=1}^n z'_j 
\prod_{i\ne j} (\wp(z_i)-\lambda)=
\sum_{k=0}^{n-1} H_k(z,z')\lambda^k .
$$
The explicit formulae for the coefficients $H_k$ are
$$
H_k=\sum_{|J|=n-k-1}(-1)^k \prod_{j\in J}\wp(z_j)\left(
\sum_{k\notin J}z'_k \right) ,
$$
where summation is taken over all subsets 
$J\subset\{1,\dots,n\}$ of cardinality $n-k-1$.

We are going to show that the functions $H_k$ are 
time-independent, \ie $d H_k/dx=0$. Indeed,
\begin{multline*}
\frac{d (-1)^k H_k(z,z')}{d x}=
\sum_{J}\prod_{j\in J}\wp(z_i)\sum_{k\notin J} z''_k+
\sum_{J}\left( \sum_{s\in J}\frac{\wp'(z_s)}{\wp(z_s)}z'_s
\right) \prod_{j\in J}\wp(z_i)\sum_{k\notin J} z''_k={}\\
{}=\sum_{J}\prod_{j\in J}\wp(z_j)\left(
\sum_{k\notin J}\sum_{i\ne k}
\frac{\wp'(z_k)+\wp'(z_i)}{\wp(z_k)-\wp(z_i)}\,z'_i z'_k 
\right)+
\sum_{J}\prod_{j\in J}\wp(z_j)\left(
\sum_{k\notin J}\sum_{s\in J}
\frac{\wp'(z_s)}{\wp(z_s)}\,z'_k z'_s\right)={}\\
{}=\sum_{J}\prod_{j\in J}\wp(z_j)\left(
\sum_{k\notin J}\sum_{s\in J}\left[
\frac{\wp'(z_s)}{\wp(z_s)}+\frac{\wp'(z_k)+\wp'(z_s)}%
{\wp(z_k)-\wp(z_s)}\right]z'_k z'_s\right)=
\sum_{J,k\notin J, s\in J} \alpha(J,k,s) ,
\end{multline*}
where
$$
\alpha(J,k,s)=\prod_{j\in J}\wp(z_j)
\left[
\frac{\wp'(z_s)}{\wp(z_s)}+\frac{\wp'(z_k)+\wp'(z_s)}%
{\wp(z_k)-\wp(z_s)}\right]z'_k z'_s  .
$$
Let us consider the involution on the set of triples
$\{J,k\notin J,s\in J\}$ which maps $\{J,k,s\}$ into $\{J',s,k\}$, 
where $J'=J\cup\{k\}\setminus\{s\}$.

Now note that
\begin{multline*}
\alpha(J,k,s)+\alpha(J',s,k)=
\prod_{j\in J\cap J'}\wp(z_j)z'_k z'_s\Bigl[
\wp'(z_s)+\wp'(z_k)\Bigr.+{}\\
{}+\left.
\wp(z_s)\frac{\wp'(z_k)+\wp'(z_s)}{\wp(z_k)-\wp(z_s)}+
\wp(z_k)\frac{\wp'(z_s)+\wp'(z_k)}{\wp(z_s)-\wp(z_k)}
\right]=0 
\end{multline*}
and therefore the whole sum 
$\sum_{J,k\notin J, s\in J} \alpha(J,k,s)$ vanishes.

\section*{Appendix II}
Consider the 2-form 
$\omega=\sum_{s=0}^{n-1} d\varphi_s\wedge dH_s$. 
Recall that $H_{s}=(-1)^s H \sigma_{n-s-1}(\wh\kappa)$,
$s=0,\dots,n-1$, where $\sigma_{n-s-1}(\wh\kappa)$ denotes 
the coefficient of $\lambda^s$ in the polynomial 
$\prod_{i=1}^{n-1}(\lambda+\wh\kappa_i)$. 
By  $\sigma_{n-s-2}^j(\wh\kappa)$ we denote the 
coefficient of $\lambda^s$ in the polynomial 
$\prod_{i\ne j}(\lambda+\wh\kappa_i)$.
Then 
\begin{multline}\label{ap2}
\omega=\sum_{s=0}^{n-1}d\varphi_s\wedge dH_s=
\sum_{s=0}^{n-2} d\varphi_s\wedge 
(-1)^s d(H\sigma_{n-s-1}(\wh\kappa))+(-1)^{n-1}\,
d\varphi_{n-1}\wedge dH={}\\
{}=\sum_{s=0}^{n-1}(-1)^s \sigma_{n-s-1}(\wh\kappa)\,
d\varphi_s\wedge dH+\sum_{j=1}^{n-1}\sum_{s=0}^{n-2}(-1)^s 
H\sigma_{n-s-2}^j(\wh\kappa)\,d\varphi_s\wedge d\wh\kappa_j .
\end{multline}
Now let us notice that
\begin{multline*}
\sum_{s=0}^{n-2}(-1)^s H\sigma_{n-s-2}^j(\wh\kappa)\,
d\varphi_s=
\sum_{s=0}^{n-2}\sum_{l=1}^n(-1)^s H
\sigma_{n-s-2}^j(\wh\kappa)\,d\intl^{\wp(z_l)}
\frac{E^s\,dE}{L(E)y(E)}={}\\
{}=\sum_{l=1}^n d\intl^{\wp(z_l)}
\frac{\suml_{s=0}^{n-2}(-1)^s H \sigma_{n-s-2}^j(\wh\kappa) 
E^s}{L(E)y(E)}\,dE -
\sum_{l=1}^n \intl^{\wp(z_l)}d\left(
\frac{\suml_{s=0}^{n-2}(-1)^s H \sigma_{n-s-2}^j(\wh\kappa) 
E^s}{L(E)y(E)}\right)\,dE={}\\
{}=\sum_{l=1}^n d\intl^{\wp(z_l)}
\frac{dE}{(E-\wh\kappa_j)y(E)}+
\sum_{l=1}^n \intl^{\wp(z_l)}
\frac{dE}{(E-\wh\kappa_j)^2 y(E)}\,d\wh\kappa_j .
\end{multline*}
In the same way one can show that
$$
\sum_{s=0}^{n-1}(-1)^s \sigma_{n-s-1}(\wh\kappa)\,d\varphi_s=
\sum_{l=1}^n d\intl^{\wp(z_l)}\frac{dE}{H y(E)}+
\sum_{l=1}^n \intl^{\wp(z_l)}\frac{dE}{H^2 y(E)}\,dH .
$$
Plugging these two formulae into (\ref{ap2}) we obtain
\begin{multline*}
d\omega=\sum_{j=1}^{n-1} d\left(\sum_{l=1}^n 
\int^{\wp(z_l)}\frac{dE}{(E-\wh\kappa_j) y(E)}\right)
\wedge d\wh\kappa_j+
d\left(\sum_{l=1}^n d\int^{\wp(z_l)}\frac{dE}{H y(E)}\right)
\wedge dH={}\\
{}=\sum_{j=1}^{n-1}d\chi_j\wedge d\wh\kappa_j+
d\chi\wedge d(\ln H) .
\end{multline*}

\subsection*{Acknowledgements}

The authors are grateful to Professor I.~M.~Krichever for
constant attention to this work.

\end{document}